\documentclass[11pt]{article}

\usepackage{jheppub}
\usepackage{amssymb,latexsym}
\usepackage{graphicx}
\usepackage{dcolumn}
\usepackage{bm}

\def \be{\begin{equation}}
\def \ee{\end{equation}}
\def \bea{\begin{eqnarray}}
\def \eea{\end{eqnarray}}
\def \nn{\nonumber}
\def \ie{{\it i.e.}}
\def \eg{{\it e.g.}}
\def \d{{\rm d}}
\def \del{\partial}

\def \snpf{S_{\rm n\hspace{-.5pt}P\hspace{-.5pt}F}}

\begin{document}

\title{Quantum Aspects of Massive Gravity II: Non-Pauli-Fierz Theory}
\author{Minjoon Park}
\affiliation{Department of Physics and MCTP, University of Michigan, \\
450 Church Street, Ann Arbor, MI 48109, USA}
\emailAdd{minjoonp@umich.edu}

\abstract{
We investigate the non-Pauli-Fierz\hspace{2pt}(nPF) theory, a linearized massive gravity with a generic graviton mass term, which has been ignored due to a ghost in its spectrum and the resultant loss of unitarity. We first show that it is possible to use the Lee-Wick mechanism, a unitarization through the decay of a ghost, in order to handle the sixth mode ghost of nPF, and then check for the quantum consistency. Once proven to be consistent, nPF could become a viable candidate for a large distance modification of gravity, because it naturally solves the intrinsic problems that most dark energy/modified gravity models suffer from: It smoothly converges to general relativity at short distances, and the small graviton mass necessary to modify gravity at large scales can be stable under the radiative corrections from the minimal gravity-to-matter coupling.
}

\maketitle

\section{Introduction}
Although the Pauli-Fierz massive gravity\cite{Fierz:1939ix} is special due to the absence of a ghost, a particle or field with a wrong-sign kinetic term, it is hard to keep it that way. While whether the Boulware-Deser ghost\cite{Boulware:1973my} can be eliminated at all nonlinear orders is still in debate\cite{Creminelli:2005qk}, we saw\cite{Park:2010rp} that quantum loop corrections revived the ghost for a large class of interactions. Then, instead of trying to obtain a ghost-free theory, how about accepting the presence of a ghost and finding a way to live with it?

Allowing a ghost to appear in the spectrum, we get the non-Pauli-Fierz\hspace{2pt}(nPF) massive gravity. Its action in a flat 4d background is
\bea\label{eqn:npfaction}
 \snpf &=& \int\d^4 x \Big\{\del_\alpha h^{\alpha\mu} \del_\beta h^\beta_\mu - \frac{1}{2}\del_\alpha h_{\mu\nu} \del^\alpha h^{\mu\nu} + \frac{1}{2} \del_\alpha h \del^\alpha h - \del_\alpha h^{\mu\alpha} \del_\mu h \nn\\
 &&\qquad\quad- \frac{m^2}{2} (h^{\mu\nu} h_{\mu\nu} - a h^2) \Big\} \,,
\eea
with $h = \eta^{\mu\nu}h_{\mu\nu}$ and $\eta_{\mu\nu} = {\rm diag}(-1,1,1,1)$. The corresponding tree level propagator is
\bea\label{eqn:npfprop}
{\bf P} &=& \frac{i}{k^2+m^2}\Big(-\frac{{\bf I}_1}{3}+\frac{{\bf I}_2}{2}+\frac{{\bf I}_3}{2m^2}-\frac{{\bf I}_4}{3m^2}+\frac{2{\bf I}_5}{4m^4}\Big) \nn\\
&&+ \frac{-i}{k^2+\frac{4a-1}{2(1-a)}m^2}\Big(\frac{{\bf I}_1}{6}-\frac{{\bf I}_4}{3m^2}+\frac{2{\bf I}_5}{3m^4}\Big)\,,
\eea
whose first line contains a massive spin-2 graviton, whereas the second line shows a spin-0 ghost. Here ${\bf I}_i$ is a set of tensor bases with 4 indicies:
\bea
{\bf I}_1 = \eta_{\mu\nu} \eta_{\lambda\rho} \,,\quad
&&{\bf I}_2 = \eta_{\mu\lambda}\eta_{\nu\rho} + \eta_{\mu\rho}\eta_{\nu\lambda} \,, \quad
{\bf I}_3 = \eta_{\mu\lambda} k_\nu k_\rho + \eta_{\mu\rho} k_\nu k_\lambda
+ (\mu \leftrightarrow \nu) \,, \nn\\
&&{\bf I}_4 = \eta_{\mu\nu} k_\lambda k_\rho + k_\mu k_\nu \eta_{\lambda\rho} \,, \quad
{\bf I}_5 = k_\mu k_\nu k_\lambda k_\rho \,.
\eea
Unlike the usual spin-projection operator of, \eg, \cite{Stelle:1976gc}, ${\bf I}_i$'s are not orthogonal, but still complete because any symmetric second rank tensor can be expanded in terms of them. When coupled to a conserved source, the $k\gg m$ limit of (\ref{eqn:npfprop}) gives the massless graviton propagator, $\frac{i}{k^2} \big(-\frac{{\bf I}_1}{2}+\frac{{\bf I}_2}{2}\big)$, of general relativity (GR) in de Donger gauge, which signifies that nPF will agree with GR at short distances. To check this claim, let us repeat van Dam and Veltman's analysis, \cite{vanDam:1970vg}, for nPF: Consider a gravitational interaction between two conserved, non-relativistic sources, $T$ and $T'$. With $g_{\rm n\hspace{-.5pt}P\hspace{-.5pt}F}$ the coupling of an nPF graviton to $T$ and $T'$, the tree level contribution to this process is
\bea
g_{\rm n\hspace{-.5pt}P\hspace{-.5pt}F}^2 T^{\mu\nu} \cdot {\bf P} \cdot T'{}^{\mu\nu} &=& g_{\rm n\hspace{-.5pt}P\hspace{-.5pt}F}^2 T^{00}\delta^\mu_0\delta^\nu_0 \Big\{\frac{i}{k^2+m^2}\Big(-\frac{{\bf I}_1}{3}+\frac{{\bf I}_2}{2}\Big) + \frac{-i}{k^2+\frac{4a-1}{2(1-a)}m^2} \frac{{\bf I}_1}{6}\Big\} T'{}^{00}\delta^\lambda_0\delta^\rho_0 \nn\\
&=& \frac{1}{2}g_{\rm n\hspace{-.5pt}P\hspace{-.5pt}F}^2 \frac{T^{00} T'{}^{00}}{k^2} + \frac{1}{k^2}{\mathcal O}\Big(\frac{m^2}{k^2}\Big) \,.
\eea
Since the same process mediated by a massless graviton is
\be
\frac{1}{2}g^2 \frac{T^{00} T'{}^{00}}{k^2}\,,
\ee
where $g$ is the gravitational coupling of GR, we see that $g_{\rm n\hspace{-.5pt}P\hspace{-.5pt}F} = g$, \ie, nPF matches GR at distances much larger than the inverse graviton mass.\footnote{In comparison, Pauli-Fierz massive gravity gives $g_{\rm P\hspace{-.5pt}F}^2 = \frac{3}{4}g^2$, which is the source of the notorious van-Dam-Veltman-Zakharov discontinuity.} Thus, if we manage to recover unitarity that is apparently violated by the ghost, nPF might become a viable candidate for a large distance modification of gravity: GR is restored at short scales, while the gravitational force gets weakened by Yukawa-suppression at distances $\gg m^{-1}$, so that long wavelength sources such as the cosmological constant gravitate less, providing a solution to the cosmological constant problem in the sense of \cite{ArkaniHamed:2002fu}.

Then, the spectrum of nPF suggests to pick up, among various proposals\cite{Hawking:2001yt} to deal with ghosts, the Lee-Wick\hspace{2pt}(LW) mechanism\cite{Lee:1969fy} as a resolution to its ghost problem. The LW mechanism unitarizes a theory with a ghost by the decay of the ghost into ordinary fields. Mainly as a means of dealing with ghosts from higher derivative theories, it has been applied to constructing a renormalizable quantum gravity\cite{Tomboulis:1977jk} as well as a hierarchy problem-free standard model\cite{Grinstein:2007mp}. To elucidate, let us highlight a scalar toy model worked out in \cite{Grinstein:2007mp}:
\be
{\mathcal L} = -\frac{1}{2}\del_\mu\sigma\del^\mu\sigma - \frac{1}{2}m_\sigma^2\sigma^2 + \frac{1}{2}\del_\mu\rho\del^\mu\rho + \frac{1}{2}m_\rho^2\rho^2 - \frac{g}{3!}(\sigma-\rho)^3\,,
\ee
where $\sigma$ is an ordinary scalar and $\rho$ is a ghost. Due to the cubic interaction, the $\rho$ propagator gets renormalized by the self-energy, $-i\Sigma$,
\be
P_\rho(p) = \frac{-i}{p^2+m_\rho^2+\Sigma(p^2)}\,.
\ee
When the $\rho\to\sigma\sigma$ decay channel is open, \ie, $m_\rho>2m_\sigma$, a contribution to $\Sigma$ from the 1-loop self-energy with two $\rho\sigma\hspace{-1pt}\sigma$-vertices has an imaginary part, $\Sigma = im_\rho\Gamma$, 
\be
\Gamma(p^2\approx-m_\rho^2) = -\frac{g^2}{32\pi m_\rho}\sqrt{1-\frac{4m_\sigma^2}{m_\rho^2}} \,< 0 \,.
\ee
Now let us consider, for example, the scattering amplitude, $\mathcal M$, of $\sigma\sigma\to\sigma\sigma$ scattering. There is a possibility of unitarity violation when the scattering is mediated by a virtual $\rho$:
\be
i{\mathcal M} = (-ig)\cdot P_\rho\cdot(-ig)\,. 
\ee
But the imaginary part of ${\mathcal M}$ for this process
\be
{\rm Im}{\mathcal M} = \frac{-g^2m_\rho\Gamma}{(p^2+m_\rho^2)^2+m_\rho^2\Gamma^2} \,,
\ee 
is still positive because the unconventional sign of $\Gamma$ is compensated by the unconventional sign of the residue of $P_\rho$, and therefore unitarity is maintained. A complete analysis of causality, stability, etc., as well as unitarity, of LW theory needs more subtle and elaborate considerations, and has been studied in the literature\cite{Cutkosky:1969fq}-\cite{Jansen:1993jj}.

It seems possible that the same mechanism would work for nPF when $\frac{4a-1}{2(1-a)}m^2 > 4m^2$. In the next section, we will see our expectation being fulfilled. Then in \S\ref{sec:renorm}, we check a basic quantum consistency of nPF. We discuss on the possible issues and future directions in \S\ref{sec:discuss}. The details of the loop computation can be found in the Appendix.

\section{LW mechanism for nPF}\label{sec:lwnpf}
Let us first see if or how the LW mechanism works for nPF. We start with (\ref{eqn:npfaction}) and decompose $h_{\mu\nu}$ in the usual way:
\be\label{eqn:hsplit}
h_{\mu\nu} = f_{\mu\nu} + \frac{\del_\mu V_\nu + \del_\nu V_\mu}{m} + \frac{\del_\mu \del_\nu \phi_1}{m^2}
+ \eta_{\mu\nu} \phi_2\,,
\ee 
with $\del^\mu f_{\mu\nu}=0$, $f=\eta^{\mu\nu}f_{\mu\nu}=0$ and $\del_\mu V^\mu=0$. Then the onshell constraint
\be
0 = \del^\nu\frac{\delta \snpf}{\delta h^{\mu\nu}} \quad\Rightarrow\quad 0 = \del^\nu h_{\mu\nu} - a\del_\mu h\,, 
\ee
gives
\be
\frac{\Box V_\mu}{m} + \frac{\del_\mu\Box\phi_1}{m^2} + \del_\mu\phi_2 = a\del_\mu\Big(\frac{\Box\phi_1}{m^2}+4\phi_2\Big)\,,
\ee
or
\be\label{eqn:vphi1}
V_\mu = 0\,,\quad \frac{\Box\phi_1}{m^2} = \frac{4a-1}{1-a}\phi_2\,.
\ee
With (\ref{eqn:hsplit}) and (\ref{eqn:vphi1}), (\ref{eqn:npfaction}) becomes
\be\label{eqn:lwaction}
 \snpf = \int \d^4x \Big(- \frac{1}{2}\del_\sigma f_{\mu\nu} \del^\sigma f^{\mu\nu} - \frac{m^2}{2} f_{\mu\nu} f^{\mu\nu} + 3 \del_\mu \phi_2 \del^\mu \phi_2 + \frac{3m^2}{2}\frac{4a-1}{1-a}\phi_2^2\Big) \,,
\ee
from which we identify the physical degrees of freedom(DOFs) to be a spin-2, $f_{\mu\nu}$, of mass $m$ and a ghost scalar, $\phi_2$, of mass $m_a=\sqrt{\frac{4a-1}{2(1-a)}}\,m$. Then we introduce a generic quantum interaction
\be\label{eqn:pni}
\int \d^4 x \, \lambda \big(\zeta_1 h^{\mu_1}_{\nu_1}h^{\nu_1}_{\mu_2}h^{\mu_2}_{\mu_1} + \zeta_2 h_{\mu\nu}h^{\mu\nu}h + \zeta_3 h^3\big)\,,
\ee
to obtain
\bea
 S &=& \int \d^4 x \Big\{-\frac{1}{4} f^{\mu\nu}(-\del^2+m^2){\bf I}_2 f^{\lambda\rho} + 3\phi_2\Big(-\del^2+\frac{4a-1}{2(1-a)}m^2\Big)\phi_2 \nn\\
 &&\qquad+ \lambda(3\zeta_1+\zeta_2) f_{\mu\nu}f^{\mu\nu}\phi_2 + \cdots\Big\}\,,
\eea
where $\cdots$ is the interactions of the form of $f^3$ and $\phi_2^3$. With
\bea
&&{\bf P}_f = \frac{i}{p^2+m^2}\frac{{\bf I}_2}{2} \equiv \includegraphics[width=1cm]{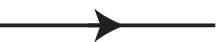} \,, \quad
P_{\phi_2} = \frac{-i}{p^2+m_a^2} \equiv \includegraphics[width=1cm]{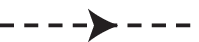} \,, \nn\\
&&{\bf V}_{ff\phi_2} = -i\lambda(3\zeta_1+\zeta_2)\frac{{\bf I}_2}{2} \equiv \includegraphics[bb=-5 10 80 50,keepaspectratio=true,width=1.3cm]{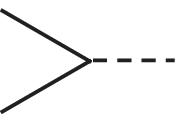} \hspace{-10pt}, \nn
\eea
we get, for $k^2 \approx -m_a^2$,
\bea\label{eqn:impart}
 -i \Sigma(k^2) &\equiv& \includegraphics[bb=0 15 75 50,keepaspectratio=true,width=1.2cm]{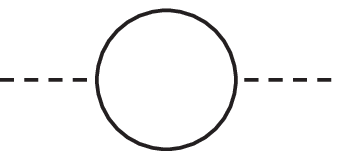}\quad = \int \frac{\d^4p}{(2\pi)^4}\big\{-i\lambda(3\zeta_1+\zeta_2)\big\}^2\frac{{\bf I}_2}{2} \frac{i}{p^2+m^2}\frac{{\bf I}_2}{2} \frac{{\bf I}_2}{2} \frac{i}{(p+k)^2+m^2}\frac{{\bf I}_2}{2} \nn\\
 &=& 5i\lambda^2(3\zeta_1+\zeta_2)^2 \Big(\frac{1}{4\pi^2\epsilon} + \frac{i}{8\pi}\sqrt{1-\frac{4m^2}{m_a^2}} + \cdots\,\Big) \,,
\eea
where we use dimensional regularization and $\cdots$ is a finite real contribution. In the narrow width approximation, (\ref{eqn:impart}) defines the decay width of $\phi_2$ into two $f$'s:
\be
P_{\phi_2} \;\Rightarrow\; P_{\phi_2,{\rm renorm}} = \frac{-i}{p^2+m_a^2+\Sigma} = \frac{-i}{p^2+m_a^2+im_a\Gamma+\cdots}\,,
\ee
with
\be
\Gamma = -\frac{5\lambda^2(3\zeta_1+\zeta_2)^2}{8\pi m_a}\sqrt{1-\frac{4m^2}{m_a^2}}\,.
\ee
Therefore, nPF theory satisfies the conditions for the LW mechanism to work when $m_a>2m$, \ie, $\frac{3}{4}<a<1$, and the imaginary part of the matrix element of the $ff\to ff$ scattering mediated by $\phi_2$,
\be
 {\rm Im}{\mathcal M} = {\rm Im}\Big[-i\{-i\lambda(3\zeta_1+\zeta_2)\}^2\frac{{\bf I}_2}{2}\cdot P_{\phi_2,{\rm renorm}}\cdot\frac{{\bf I}_2}{2}\Big] \propto \frac{-m_a\Gamma}{(p^2+m_a^2)^2+m_a^2\Gamma^2}\,,
\ee
is positive, as required by unitarity.

\section{Renormalizability}\label{sec:renorm}
Since quantum effects play an essential role in rendering nPF theory sensible, we have to check the quantum consistency of the theory. One of the items in the checklist is renormalizability. Divergences of the loop corrections to $\snpf$ would have pieces of ${\mathcal O}(k^4)$ and higher orders as well as ${\mathcal O}(k^0)$ and ${\mathcal O}(k^2)$ terms. Adopting the idea\cite{Donoghue:1994dn} of effective field theory (EFT), we may assume those divergences of higher orders in $k$ can be absorbed into EFT terms. But regardless of the EFT treatment, ${\mathcal O}(k^0)$ and ${\mathcal O}(k^2)$ divergences must be taken care of at tree level. With the divergences of the 1PI 2-point function written as
\be\label{eqn:2p1pi}
 {\bf \Pi}_{\rm div} = \frac{1}{\epsilon}\big\{b_{1,m}{\bf I}_1 + b_{2,m}{\bf I}_2 + k^2\big(b_{1,k}{\bf I}_1 + b_{2,k}{\bf I}_2\big) + b_3{\bf I}_3 + b_4{\bf I}_4 + {\mathcal O}(k^4)\big\} \,,
\ee
the very first requirement for any desired interaction is that its ${\mathcal O}(k^2)$ part conform to the tree level form, $k^2\big({\bf I}_1-\frac{{\bf I}_2}{2}\big)+\frac{{\bf I}_3}{2}-{\bf I}_4$, \ie,
\be\label{eqn:k2req}
b_{1,k} + 2b_{2,k}=0\,,\;\; b_{1,k}+b_4=0\,,\;\; b_{2,k}+b_3=0\,.
\ee
When there is a 1PI 1-point function(tadpole), 
\be
{\bf B}_{\rm div} = \frac{1}{\epsilon} b_0 \eta_{\mu\nu}\,,
\ee
it should be interpreted as the ${\mathcal O}(h)$ piece of $\Lambda\sqrt{-{\rm det}\big(\eta_{\mu\nu}+\frac{\hat h_{\mu\nu}}{M_P}\big)}=\Lambda\big(1+\frac{\hat h}{2M_P}+\frac{\hat h^2}{8M_P^2}-\frac{\hat h_{\mu\nu}\hat h^{\mu\nu}}{4M_P^2}+\cdots\big)$. Then the conformity at ${\mathcal O}(k^0)$ means
\be
b_0 \hat h + \hat h\cdot(b_{1,m}{\bf I}_1+b_{2,m}{\bf I}_2)\cdot \hat h
\ee 
should be able to be written as
\be
\Lambda\Big\{\frac{\hat h}{2M_P}+\frac{\hat h\hat h}{8M_P^2}({\bf I}_1-{\bf I}_2)\Big\} + X\hat h\hat h\Big(-a{\bf I}_1+\frac{{\bf I}_2}{2}\Big)\,,
\ee
for some $\Lambda$ and $X$. That is, the second requirement is
\be\label{eqn:k0req}
b_{1,m}+2a b_{2,m} = (1-2a)\frac{b_0}{4M_P}\,.
\ee

\subsection{Polynomial interaction}
At 1-loop, the ${\mathcal O}(k^2)$ part of (\ref{eqn:2p1pi}) for the cubic polynomial interaction, (\ref{eqn:pni}), is calculated in \ref{sec:pniloop}, from which we can obtain the explicit expression for (\ref{eqn:k2req}):
\bea\label{eqn:k2reqpni}
0 &=& 9(37-74a+52a^2+84a^3)\zeta_1^2 \nn\\
&&+ 4\zeta_2\big\{(193-462a+512a^2+144a^3)\zeta_2+144(6a-1)\zeta_3\big\} \nn\\
&&+ 12\zeta_1\big\{(81-190a+200a^2+80a^3)\zeta_2+24(7a-1)\zeta_3\big\} \,, \nn\\
0 &=& 9(61-162a+100a^2+52a^3)\zeta_1^2 \\
&&+ 4\zeta_2\big\{(185-414a+352a^2+144a^3)\zeta_2+288(3a-1)\zeta_3\big\} \nn\\
&&+ 12\zeta_1\big\{(125-322a+256a^2+64a^3)\zeta_2+12(22a-7)\zeta_3\big\} \,, \nn\\
0 &=& 3(7-16a+16a^2-4a^3)\zeta_1^2 + 4(2a-1)\zeta_2^2 + 4\zeta_1\zeta_2(2a+1) \,. \nn
\eea
Unfortunately, the only real set of $(\zeta_1,\zeta_2,\zeta_3,a)$ that solves (\ref{eqn:k2reqpni}) is
\be
\zeta_2=-1.03\zeta_1\,,\;\;\zeta_3=2.01\zeta_1\,,\;\;a=2.65\,,
\ee
and we get $m_a^2<0$, \ie, a tachyon instability. Therefore nPF theory with a generic polynomial 3-graviton interaction is inconsistent either due to nonrenormalizability or instability.

\subsection{Derivative + polynomial interaction}
A more general interaction may contain derivatives, and a reasonable way to introduce such interactions is to expand $\sqrt{-g}R$ to higher orders in $h$ with $g = \eta+\frac{h}{M_P}$. That is, we now investigate
\bea\label{eqn:4thonpfa}
 S &=& \snpf 
+ 2M_P^2\int\d^4x \,\Big\{ [ \sqrt{-g}\,R\,]\big|_{h^3} + [ \sqrt{-g}\,R\,]\big|_{h^4} \nn\\
 &&- \frac{m^2}{4M_P^3}\big( x_1 h^\mu_\nu h^\nu_\sigma h^\sigma_\mu + x_2 h_{\mu\nu}h^{\mu\nu}h
+ x_3 h^3 \big) \nn\\
 &&- \frac{m^2}{4M_P^4}\big( y_1 h^\mu_\nu h^\nu_\sigma h^\sigma_\lambda h^\lambda_\mu 
+ y_2 (h_{\mu\nu}h^{\mu\nu})^2 + y_3 h^\mu_\nu h^\nu_\sigma h^\sigma_\mu h
+ y_4 h_{\mu\nu}h^{\mu\nu} h^2 + y_5 h^4 \big) \Big\} \nn\\
 &=& \snpf + \int \d^4x\, \big(i{\bf V}_{3h}hhh + i{\bf V}_{4h}hhhh\big) \,.
\eea
Now that $M_P^{-1}$ plays the role of the coupling, we need the cubic and quartic vertices in order to get loops of ${\mathcal O}(M_P^{-2})$. Explicit form of $\sqrt{-g}\,R\big|_{h^3}$ and $\sqrt{-g}\,R\big|_{h^4}$ can be found in, \eg, \cite{gvertices}, and ${\bf V}_{3h}$ and ${\bf V}_{4h}$ as well as ingredients of the 1PI 1-point and 2-point functions are given in \ref{sec:der+pniloop}. Then the condition (\ref{eqn:k2req}) is
\bea
 0 &=& 536+333x_1^2+36x_1(54+27x_2-8x_3)+4x_2(402+193x_2-144x_3)+1032x_3 \nn\\
 &&- a\big\{2104+999x_1^2+4x_2(1664+655x_2) \nn\\
 &&\qquad+ 12x_1(638+271x_2-192x_3)+48(97-84x_2)x_3\big\} \nn\\
 &&+ 2a^2\big\{1586+5694x_1+567x_1^2+5092x_2+2340x_1x_2 \nn\\
 &&\qquad+ 1948x_2^2+48(58-21x_1-36x_2)x_3\big\} \nn\\
 &&- 4a^3\big\{559-72x_1^2+30x_1(65+12x_2)+4x_2(439+92x_2)\big\} \nn\\
 &&+ a^4\big\{632-756x_1^2+48x_1(51-20x_2)+64x_2(43-9x_2)\big\}  \,, \nn\\
 0 &=& 138+549x_1^2+6x_1(145+250x_2-168x_3)+740x_2^2+16x_2(67-72x_3)+156x_3 \nn\\
 &&- a\big\{598+2007x_1^2+2396x_2^2+12x_1+960x_3+48x_2(95-96x_3)\big\} \\
 &&+ 2a^2\big\{556+1179x_1^2+6x_1(433+578x_2-264x_3) \nn\\
 &&\qquad+ 4x_2(832+383x_2-432x_3)+888x_3\big\} \nn\\
 &&- 4a^3\big\{253+108x_1^2+4x_2(233+52x_2)+6x_1(131+96x_2)\big\} \nn\\
 &&+ 4a^4\big\{90+248x_2+3x_1(68-39x_1)-48x_2(4x_1+3x_2)\big\} \,, \nn\\
 0 &=& 28+63x_1^2+4x_2(2-3x_2)+12x_1(7+x_2) \nn\\
 &&- a\big\{12(2+3x_1)(3+4x_1)-8x_2(5+3x_1)-24x_2^2\big\} \nn\\
 &&+ 12a^2(3+14x_1+12x_1^2-4x_2) + a^3(8-48x_1-36x_1^2) \,, \nn
\eea
which admits numerous solutions, such as
\be\label{eqn:xasol}
x_1=1.00\,,\;\; x_2=-4.06\,,\;\; x_3=1.68\,,\;\; a=0.85\,.
\ee
For this choice of $x$'s and $a$, the tadpole condition (\ref{eqn:k0req}),
\bea
 0 &=& 65+3x_1(191+204x_1+96x_2-2808x_3)+x_2(822-816x_2-12528x_3) \nn\\
 &&\qquad- 108x_3(23+36x_3) + 216y_1-144y_2+1512y_3+2304y_4+1728y_5 \nn\\
 &&- a\big\{472+3x_1(1483+2442x_1+4968x_2-10368x_3) \nn\\
 &&\qquad+2x_2(5145+2988x_2-25920x_3)-108x_3(73+288x_3) \nn\\
 &&\qquad- 408y_1-6000y_2+12096y_3+18720y_4+15552y_5\big\} \nn\\
 &&+ a^2\big\{1273+12x_1(751+2010x_1+4974x_2-2916x_3) \nn\\
 &&\qquad+ 12x_2(3029+2948x_2-5616x_3)-648x_3(11+96x_3) \nn\\
 &&\qquad- 12624y_1-43104y_2+33048y_3+50976y_4+41472y_5\big\} \nn\\
 &&- 2a^3\big\{1283+6x_1(355+2946x_1+7512x_2-864x_3) \\
 &&\qquad+ 24x_2(1243+1258x_2-288x_3)-1728x_3 \nn\\
 &&\qquad- 21552y_1-59520y_2+21168y_3+30528y_4+13824y_5\big\} \nn\\
 &&+ 4a^4\big\{1111-3x_1(443-2307x_1-4896x_2+1296x_3) \nn\\
 &&\qquad+ 12x_2(1093+780x_2-432x_3)-432x_3 \nn\\
 &&\qquad- 15600y_1-39840y_2+7560y_3+10080y_4\big\} \nn\\
 &&- 8a^5\big\{539-x_1(903-1908x_1-3168x_2)+12x_2(241+160x_2) \nn\\
 &&\qquad- 5736y_1-13920y_2+1296y_3+1728y_4\big\} \nn\\
 &&+ 32a^6\big\{49-3x_1(29-12x_1)+108x_2-456y_1-1056y_2\big\} \,,\nn
\eea
becomes
\be
0=-3407+197y_1+398y_2+293y_3+639y_4+1492y_5\,,
\ee
and it can be solved for various choices of $y$'s. Since the $\phi_2\to ff$ decay channel is still available in (\ref{eqn:4thonpfa}) and $a$ from (\ref{eqn:xasol}) lies within $\big(\frac{3}{4},1\big)$, the LW mechanism is applicable. Therefore, nPF with both derivative and polynomial interactions may give a unitary and consistent theory of an effective quantum massive gravity.

\section{Discussion}\label{sec:discuss}
The solidity of the LW mechanism has not been fully established yet. Although there is no known example against the LW mechanism, and the consistency of several specific models, \eg, \cite{Jansen:1993jj}, was thoroughly verified, other cases, including nPF, need to be worked out in this respect.

Once its consistency is checked, nPF might deserve more attention, because it has advantages over other dark energy or modified gravity models.
\begin{itemize}
\item Short distance behavior: To modify gravity at large distances, an extra scalar DOF is a requisite. But once incorporated into the gravity sector, the universal coupling of this scalar to sources persists at all scales, ruining the matching of the theory to the observations and experiments at the solar system scale and below. Therefore, such models need to invoke external mechanisms such as Vainshtein\cite{vainshtein} or chameleon\cite{Khoury:2003aq}, to make themselves viable. But as mentioned in the introduction, we can obtain a smooth limit to GR automatically in nPF: Although nPF has two extra scalars, one normal and the other ghost, at distances much smaller than $m^{-1}$ the effects coming from the coupling of the ghost to matter cancel those from the normal scalar-to-matter coupling.
\item Radiative stability: Generically any dark energy or modified gravity model requires a small scale to be introduced to explain the cosmic acceleration at the Hubble scale. Keeping such a small scale safe from radiative corrections from, \eg, the Standard Model fields is nontrivial and sometimes needs a fine-tuning at the same level of the cosmological constant problem. On the other hand, the loop corrections\cite{Park:2010rp} to any gravity theory from the Standard Model fields minimally coupled to the graviton take the form of 
\be\label{eqn:mea}
\sim \int \d^4x \sqrt{-g} (\Lambda + a_1 R + a_2 R^2 + a_3 R_{\mu\nu}R^{\mu\nu} + \cdots)\,.
\ee
Then the renormalized cosmological constant can get large due to $\Lambda$ in (\ref{eqn:mea}), but in a massive gravity theory the Yukawa suppressed gravitational force dilutes it at large distances, solving the cosmological constant problem. At the same time, (\ref{eqn:mea}) does not contribute to the nPF graviton mass term
\be
\int \d^4x \sqrt{-g} \Big(-\frac{m^2}{2}\Big) (h_{\mu\nu}h^{\mu\nu} - a h^2)\,,
\ee 
for $a\neq\frac{1}{2}$, because the contribution to $h_{\mu\nu}h^{\mu\nu}$ or $h^2$ from the perturbative expansion of (\ref{eqn:mea}) is only
\be
\int \d^4x \sqrt{-g} \,\Lambda (h_{\mu\nu}h^{\mu\nu} - \frac{1}{2}h^2)\,.
\ee
In order for the LW mechanism to work, we require $\frac{3}{4}<a<1$, so that $m$ does not get affected by (\ref{eqn:mea}). In other words, since the matter loop corrections should be covariant, they cannot renormalize the non-covariant entity, such as the graviton mass, so that $m$ is radiatively stable under the minimal gravity-to-matter coupling. But we cannot claim complete radiative stability of $m$ yet: We presented nPF only in a perturbative form, and a consistent, nonlinear completion of nPF may require a certain nonminimal coupling of gravity to matter, with which $m$ could get radiative corrections. 
\end{itemize}
With these benefits and peculiarities, nPF might be able to play a role not only in cosmology, but in solar system anomalies such as the Pioneer anomaly and even in the Standard Model--gravity interactions. 

The analysis so far is performed on a Minkowski background as a proof of concept. But in order to apply it to cosmology, we need to generalize it to curved backgrounds. This generalization seems straightforward, although the actual algebra gets complicated since we can no longer work in the 4-momentum space; The theory will be defined with the covariantized version of (\ref{eqn:npfaction}), and we may expect that the decay of the ghost through the LW mechanism will be seen by an imaginary shift of the ghost mass due to renormalization. There is also an advantage in working in a curved spacetime. For example, if we work in (Anti-)de Sitter background, a bare theory has one more parameter, the (A)dS curvature, so that the requirements for renormalizability, (\ref{eqn:k2req}) and (\ref{eqn:k0req}), may be relaxed. But how the LW mechanism will work in a curved background needs more investigation, because there are subtleties in quantum field theory in curved spacetimes, \eg, the preparation of a vacuum. It would also be interesting to look into whether nPF can evade generic problems of Pauli-Fierz massive gravity in curved backgrounds, such as the Higuchi ghost\cite{Higuchi:1986py} or some issues raised in \cite{Grisa:2009yy}.

\acknowledgments
We thank Ratindranath Akhoury, Ian-Woo Kim, James Liu and Lorenzo Sorbo for fruitful discussions.

\appendix
\section{nPF massive graviton loops}
We can add arbitrary cubic and quartic interactions to $\snpf$ by introducing the cubic and quartic vertices, ${\bf V}_{3h}$ and ${\bf V}_{4h}$:
\be
S = \snpf + \int \d^4x\, \big(i{\bf V}_{3h}hhh + i{\bf V}_{4h}hhhh\big)\,.
\ee
To find loop corrections to $\snpf$, we path-integrate over $h$, with ${\cal O}(h^3)$ and ${\cal O}(h^4)$ terms providing quantum interactions:
\bea\label{eqn:npfpf}
 Z[J] &=& \int {\cal D}h\, \exp\Big[\,i \Big(S + \int \d^4 x\, J^{\mu\nu}h_{\mu\nu} \Big) \Big] \nn\\
 &=& {\cal N} \exp\Big[\, i \int\d^4 z \Big\{i{\bf V}_{3h}\Big(\frac{\delta}{i\delta J(z)}\Big)^3 + i{\bf V}_{4h}\Big(\frac{\delta}{i\delta J(z)}\Big)^4 \Big\} \Big] \nn\\
&&\qquad\times\exp\Big[\, \frac{i}{2}\int \d^4 x \d^4 y J(x) \big(-i\tilde{\bf P}(x-y)\big) J(y) \Big]\,,
\eea
where ${\cal N}$ is a normalization constant and $\tilde{\bf P}$ is the inverse Fourier transform of the tree level nPF graviton propagator, (\ref{eqn:npfprop}). With 
\be
\tilde{\bf P} = \includegraphics[width=1cm]{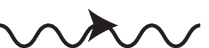} \,,
\quad
i\int\d^4x J(x) \big(-i\tilde{\bf P}(x-y)\big) = \includegraphics[width=1cm]{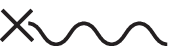} \,,
\ee
we can expand (\ref{eqn:npfpf}) diagrammatically. The pieces of our interest are
\vspace{5pt}
\bea\label{eqn:lmgi}
&& -3i \includegraphics[bb=-5 15 70 50,keepaspectratio=true,width=1.6cm]{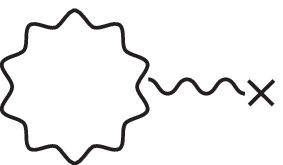}
\quad - 6 \includegraphics[bb=-5 15 70 50,keepaspectratio=true,width=1.6cm]{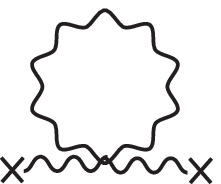}
- \frac{1}{2} \Big( 18 \includegraphics[bb=-5 15 70 50,keepaspectratio=true,width=1.6cm]{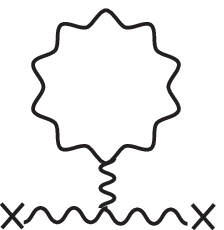} 
+ 18 \includegraphics[bb=-5 15 70 50,keepaspectratio=true,width=1.6cm]{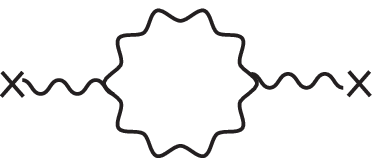} \qquad\; \Big) \,, \\
&&\hspace{24pt}\underbrace{\hspace{20pt}}_{(a)} \hspace{59pt}\underbrace{\hspace{20pt}}_{(b)} \hspace{62pt}\underbrace{\hspace{20pt}}_{(c)} \hspace{61pt}\underbrace{\hspace{20pt}}_{(d)} \nn
\eea
where $(a)\sim(d)$ are the loop parts of corresponding diagrams.

\subsection{Polynomial interaction}\label{sec:pniloop}
For the interaction of (\ref{eqn:pni}), there is only the 3-point vertex
\bea\label{eqn:3pv}
 {\bf V}_{3h} =& -\frac{i\lambda}{(2!)^33!} \big( \zeta_1 \eta_{\mu_1\nu_3} \eta_{\nu_1\mu_2} \eta_{\nu_2\mu_3} + \zeta_2 \eta_{\mu_1\mu_2} \eta_{\nu_1\nu_2} \eta_{\mu_3\nu_3} + \zeta_3 \eta_{\mu_1\nu_1} \eta_{\mu_2\nu_2} \eta_{\mu_3\nu_3} \nn\\
 &\hspace{50pt} + {\rm symmetrization\;in\;}\mu\nu + {\rm permutation\;in\;}123 \big) \,,
\eea
and we have only to consider diagrams $(a)$, $(c)$ and $(d)$. Among their divergences, only that of $(d)$ has ${\mathcal O}(k^2)$ pieces:
\bea
 && \frac{1}{\epsilon}\frac{i \lambda^2}{6912\pi^2m^2(1-a)^3} \Big[\big\{27(9-30a+28a^2+4a^3)\zeta_1^2 \nn\\
 &&\qquad+ 4\zeta_2\big((179-466a+512a^2+144a^3)\zeta_2+144(6a-1)\zeta_3\big) \nn\\
 &&\qquad+ 12\zeta_1\big((67-186a+168a^2+80a^3)\zeta_2+24(7a-1)\zeta_3\big)\big\}k^2\frac{{\bf I}_1}{2} \nn\\
 &&\quad+ \big\{9(5+8a-16a^2+36a^3)\zeta_1^2+12(7-2a+16a^2)\zeta_1\zeta_2+4(7+2a)\zeta_2^2\big\}k^2\frac{{\bf I}_2}{2} \nn\\
 &&\quad+ \big\{9(15-44a+48a^2-28a^3)\zeta_1^2-12(1-6a+8a^2)\zeta_1\zeta_2-4(11-14a)\zeta_2^2  \big\}{\bf I}_3 \nn\\
 &&\quad+ \big\{9(17-36a+8a^2+20a^3)\zeta_1^2 + 4\zeta_2\big((3+26a-80a^2)\zeta_2-72\zeta_3\big) \nn\\
 &&\qquad+ 12\zeta_1\big((29-68a+44a^2-8a^3)\zeta_2+6(8a-5)\zeta_3\big)\big\}{\bf I}_4 \Big] \,,
\eea
from which we can read off $b_{1,k}$, $b_{2,k}$, $b_3$ and $b_4$.

\subsection{Derivative + polynomial interaction}\label{sec:der+pniloop}
For the more complicated interaction of (\ref{eqn:4thonpfa}), we have
\bea
 &&{\bf V}_{3h} = -\frac{i}{M_P}\Big\{
-\frac{1}{4}k_1\cdot k_2\eta_{\mu_1\nu_1}\eta_{\mu_2\nu_2}\eta_{\mu_3\nu_3} 
- \frac{1}{2}k_{1\mu_2}k_{1\nu_2}\eta_{\mu_1\nu_1}\eta_{\mu_3\nu_3} \nn\\
 &&\quad+ \frac{1}{4}k_1\cdot k_2\eta_{\mu_1\mu_2}\eta_{\nu_1\nu_2}\eta_{\mu_3\nu_3} 
+ k_1\cdot k_2\eta_{\mu_1\nu_1}\eta_{\mu_2\mu_3}\eta_{\nu_2\nu_3} 
+ k_{1\mu_2}k_{1\nu_3}\eta_{\mu_1\nu_1}\eta_{\nu_2\mu_3} \nn\\
 &&\quad- \frac{1}{2}k_{1\nu_2}k_{2\mu_1}\eta_{\nu_1\mu_2}\eta_{\mu_3\nu_3} 
+ \frac{1}{2}k_{1\mu_3}k_{2\nu_3}\eta_{\mu_1\mu_2}\eta_{\nu_1\nu_2} 
+ k_{1\mu_3}k_{1\nu_3}\eta_{\mu_1\mu_2}\eta_{\nu_1\nu_2} \\
 &&\quad+ 2k_{1\mu_2}k_{2\nu_3}\eta_{\mu_1\nu_2}\eta_{\nu_1\mu_3} 
+ k_{1\mu_2}k_{2\mu_1}\eta_{\nu_1\nu_3}\eta_{\nu_2\mu_3} 
- k_1\cdot k_2\eta_{\nu_1\mu_2}\eta_{\nu_2\mu_3}\eta_{\nu_3\mu_1} \nn\\
 &&\quad- \frac{m^2}{2}\big(x_1\eta_{\mu_1\mu_2}\eta_{\nu_1\mu_3}\eta_{\nu_2\nu_3} 
+ x_2\eta_{\mu_1\mu_2}\eta_{\nu_1\nu_2}\eta_{\mu_3\nu_3} 
+ x_3\eta_{\mu_1\nu_1}\eta_{\mu_2\nu_2}\eta_{\mu_3\nu_3}\big) \nn\\
 &&\quad+ {\rm symmetrization\;in\;}\mu\nu + {\rm permutation\;in\;}123 \Big\} \,, \nn\\
 &&{\bf V}_{4h} = -\frac{i}{M_P^2}\Big\{
-\frac{1}{16}k_1\cdot k_2\eta_{\mu_1\nu_1}\eta_{\mu_2\nu_2}\eta_{\mu_3\nu_3}\eta_{\mu_4\nu_4}
- \frac{1}{8}k_{1\mu_2}k_{1\nu_2}\eta_{\mu_1\nu_1}\eta_{\mu_3\nu_3}\eta_{\mu_4\nu_4} \nn\\
 &&\quad- \frac{1}{8}k_{1\mu_2}k_{2\mu_1}\eta_{\nu_1\nu_2}\eta_{\mu_3\nu_3}\eta_{\mu_4\nu_4}
+ \frac{1}{16}k_1\cdot k_2\eta_{\mu_1\mu_2}\eta_{\nu_1\nu_2}\eta_{\mu_3\nu_3}\eta_{\mu_4\nu_4} \nn\\
 &&\quad+ \frac{1}{8}k_1\cdot k_2\eta_{\mu_1\nu_1}\eta_{\mu_2\nu_2}\eta_{\mu_3\mu_4}\eta_{\nu_3\nu_4}
+ \frac{1}{4}k_{1\mu_2}k_{1\nu_2}\eta_{\mu_1\nu_1}\eta_{\mu_3\mu_4}\eta_{\nu_3\nu_4} \nn\\
 &&\quad+ \frac{1}{4}k_{1\mu_2}k_{2\mu_1}\eta_{\nu_1\nu_2}\eta_{\mu_3\mu_4}\eta_{\nu_3\nu_4}
- \frac{1}{8}k_1\cdot k_2\eta_{\mu_1\mu_2}\eta_{\nu_1\nu_2}\eta_{\mu_3\mu_4}\eta_{\nu_3\nu_4}\nn\\
 &&\quad+ \frac{1}{2}k_1\cdot k_2\eta_{\mu_1\nu_1}\eta_{\mu_2\mu_3}\eta_{\nu_2\nu_3}\eta_{\mu_4\nu_4}
+ \frac{1}{2}k_{1\mu_2}k_{1\nu_2}\eta_{\mu_1\mu_3}\eta_{\nu_1\nu_3}\eta_{\mu_4\nu_4}\nn\\
 &&\quad+ \frac{1}{4}k_{1\mu_3}k_{2\nu_3}\eta_{\mu_1\mu_2}\eta_{\nu_1\nu_2}\eta_{\mu_4\nu_4}
- \frac{1}{2}k_1\cdot k_2\eta_{\nu_1\mu_2}\eta_{\nu_2\mu_3}\eta_{\nu_3\mu_1}\eta_{\mu_4\nu_4} \nn\\
 &&\quad+ \frac{1}{2}k_{1\mu_2}k_{2\mu_1}\eta_{\nu_2\mu_3}\eta_{\nu_3\nu_1}\eta_{\mu_4\nu_4}
+ \frac{1}{2}k_{1\mu_2}k_{1\mu_3}\eta_{\nu_2\nu_3}\eta_{\mu_1\nu_1}\eta_{\mu_4\nu_4} \\
 &&\quad- k_1\cdot k_2\eta_{\mu_1\nu_1}\eta_{\nu_2\mu_3}\eta_{\nu_3\mu_4}\eta_{\nu_4\mu_2}
- 2k_{1\mu_2}k_{2\mu_3}\eta_{\nu_2\mu_4}\eta_{\nu_4\mu_1}\eta_{\nu_1\nu_3} \nn\\
 &&\quad+ \frac{1}{2}k_1\cdot k_2\eta_{\nu_1\mu_3}\eta_{\nu_3\mu_2}\eta_{\nu_2\mu_4}\eta_{\nu_4\mu_1}
- \frac{1}{2}k_1\cdot k_2\eta_{\mu_1\mu_3}\eta_{\nu_1\nu_3}\eta_{\mu_2\mu_4}\eta_{\nu_2\nu_4} \nn\\
 &&\quad- \frac{1}{2}k_{1\mu_3}k_{2\mu_4}\eta_{\nu_3\nu_4}\eta_{\mu_1\mu_2}\eta_{\nu_1\nu_2}
- 2k_{1\mu_2}k_{2\mu_3}\eta_{\nu_2\mu_1}\eta_{\nu_1\mu_4}\eta_{\nu_4\nu_3}\nn\\
 &&\quad+ k_1\cdot k_2\eta_{\nu_1\mu_2}\eta_{\nu_2\mu_3}\eta_{\nu_3\mu_4}\eta_{\nu_4\mu_1} 
- k_{1\mu_2}k_{1\nu_2}\eta_{\nu_1\mu_3}\eta_{\nu_3\mu_4}\eta_{\nu_4\mu_1} \nn\\
 &&\quad- k_{1\mu_3}k_{2\nu_3}\eta_{\nu_1\mu_4}\eta_{\nu_4\mu_2}\eta_{\nu_2\mu_1}
+ k_{1\mu_2}k_{2\mu_3}\eta_{\nu_2\mu_1}\eta_{\nu_1\nu_3}\eta_{\mu_4\nu_4}
- k_{1\mu_3}k_{2\mu_4}\eta_{\nu_3\mu_2}\eta_{\nu_2\mu_1}\eta_{\nu_1\nu_4} \nn\\
 &&\quad- k_{1\mu_2}k_{1\mu_3}\eta_{\mu_1\nu_1}\eta_{\nu_2\mu_4}\eta_{\nu_4\nu_3}
- k_{1\mu_2}k_{1\mu_3}\eta_{\nu_2\nu_3}\eta_{\mu_1\mu_4}\eta_{\nu_1\nu_4}
- k_{1\mu_2}k_{2\mu_1}\eta_{\nu_2\mu_3}\eta_{\nu_3\mu_4}\eta_{\nu_4\nu_1} \nn\\
 &&\quad- \frac{m^2}{2}\big(y_1\eta_{\mu_1\mu_2}\eta_{\nu_1\mu_3}\eta_{\nu_2\mu_4}\eta_{\nu_3\nu_4}
+ y_2\eta_{\mu_1\mu_2}\eta_{\nu_1\nu_2}\eta_{\mu_3\mu_4}\eta_{\nu_3\nu_4}
+ y_3\eta_{\mu_1\mu_2}\eta_{\nu_1\mu_3}\eta_{\nu_2\nu_3}\eta_{\mu_4\nu_4}\nn\\
 &&\qquad+ y_4\eta_{\mu_1\mu_2}\eta_{\nu_1\nu_2}\eta_{\mu_3\nu_3}\eta_{\mu_4\nu_4}
+ y_5\eta_{\mu_1\nu_1}\eta_{\mu_2\nu_2}\eta_{\mu_3\nu_3}\eta_{\mu_4\nu_4}\big) \nn\\
 &&\quad+ {\rm symmetrization\;in\;}\mu\nu + {\rm permutation\;in\;}1234 \Big\} \,. \nn
\eea
Then the divergences up to ${\mathcal O}(k^4)$ are 
\bea
 && (a) = \frac{1}{\epsilon}\frac{im^4}{256\pi^2M_P(1-a)^3}\big\{7+21x_1+30x_2+12x_3-a(23+66x_1+96x_2+48x_3) \nn\\
 &&\quad+ 4a^2(7+18x_1+24x_2)-12a^3(1+3x_1+4x_2)\big\}\eta_{\mu\nu} + {\rm finite} \,,\\
 && (b) = \frac{1}{\epsilon}\frac{i m^2}{512\pi^2M_P^2(1-a)^3}\Big[(-7+22a-24a^2+12a^3)\Big(k^2{\bf I}_1-k^2\frac{{\bf I}_2}{2}+\frac{{\bf I}_3}{2}-{\bf I}_4\Big) \nn\\
 &&\quad+ \frac{m^2}{3}\big\{3(3y_1-2y_2+21y_3+32y_4+24y_5)-2a(14y_1-8y_2+99y_3+156y_4+144y_5) \nn\\
 &&\qquad+ 4a^2(2-a)(4y_1-4y_2+27y_3+36y_4)\big\}{\bf I}_1 \nn\\
 &&\quad- \frac{m^2}{6}\big\{3(7-30y_1-68y_2-3y_3-4y_4)-a(69-284y_1-640y_2-36y_3-48y_4) \nn\\
 &&\qquad+ 4a^2(21-76y_1-176y_2)-4a^3(9-38y_1-88y_2)\big\}{\bf I}_2 + {\mathcal O}(k^4)\Big] + {\rm finite} \,,\\
 && (c) = \frac{1}{\epsilon}\frac{i m^2}{1536\pi^2M_P^2(1-a)^3(4a-1)}\big\{7+21x_1+30x_2+12x_3-a(23+66x_1+96x_2+48x_3) \nn\\
 &&\qquad\quad+ 4a^2(7+18x_1+24x_2)-12a^3(1+3x_1+4x_2)\big\} \nn\\
 &&\qquad \Big[\Big(k^2{\bf I}_1-k^2\frac{{\bf I}_2}{2}+\frac{{\bf I}_3}{2}-{\bf I}_4\Big) + 2m^2(x_2+6x_3){\bf I}_1 + \frac{m^2}{2}(3x_1+4x_2){\bf I}_2 + {\mathcal O}(k^4)\Big] + {\rm finite} \,,\nn\\
\eea
\bea
 && (d) = \frac{1}{\epsilon}\frac{i m^2}{55296\pi^2M_P^2(1-a)^4}\Big[
\big\{462+243x_1^2+12x_1(33+67x_2-24x_3) \nn\\
 &&\qquad\quad+ 2(792x_2+358x_2^2+516x_3-288x_2x_3) \nn\\
 &&\qquad- a\big(1978+1053x_1^2+6304x_2+2580x_2^2 \nn\\
 &&\qquad\quad+ 12x_1(138+253x_2-192x_3)+4656x_3-4032x_2x_3\big) \nn\\
 &&\qquad+ 2a^2\big(1592+783x_1^2+6x_1(233+354x_2-168x_3) \nn\\
 &&\qquad\quad+ 4(1167x_2+489x_2^2+696x_3-432x_2x_3)\big) \nn\\
 &&\qquad- 4a^3\big(623+162x_1^2+1572x_2+368x_2^2+6x_1(107+44x_2)\big) \nn\\
 &&\qquad+ 4a^4\big(206-27x_1^2+688x_2-144x_2^2+60x_1(7-4x_2)\big)\big\}k^2{\bf I}_1 \nn\\
 &&\quad+\big\{37+774x_1+45x_1^2+12x_2+84x_1x_2+28x_2^2 \nn\\
 &&\qquad- a\big(63-27x_1^2+176x_2+20x_2^2+12x_1(250+9x_2)\big) \nn\\
 &&\qquad- a^2\big(6+216x_1^2-424x_2+8x_2^2-24x_1(179+9x_2)\big) \nn\\
 &&\qquad+ 4a^3\big(32+117x_1^2-92x_2-6x_1(109+8x_2)\big) - 12a^4(8-32x_1+27x_1^2)\big\}k^2{\bf I}_2 \nn\\
 &&\quad+\big\{103+270x_1^2+28x_2-88x_2^2-6x_1(59+4x_2) \nn\\
 &&\qquad- a\big(437+1062x_1^2-336x_2-200x_2^2-24x_1(65+7x_2)\big) \nn\\
 &&\qquad+ 2a^2\big(273+828x_1^2-432x_2-56x_2^2-42x_1(29+4x_2)\big) \nn\\
 &&\qquad- 4a^3\big(67+342x_1^2-152x_2-48x_1(8+x_2)\big) + 8a^4(7-18x_1+63x_1^2)\big\}{\bf I}_3 \\
 &&\quad- 2\big\{162-153x_1^2+256x_2-12x_2^2-3x_1(79+116x_2-120x_3)+438x_3+288x_2x_3 \nn\\
 &&\qquad- a\big(690-477x_1^2-12x_1(80+97x_2-78x_3)+4(218x_2+23x_2^2+462x_3+72x_2x_3)\big) \nn\\
 &&\qquad+ 4a^2\big(259-99x_1^2+335x_2+106x_2^2-12x_1(25+28x_2-12x_3)+474x_3\big) \nn\\
 &&\qquad- 4a^3\big(185+27x_1^2-12x_1(6+13x_2)+80x_2(4+x_2)\big) \nn\\
 &&\qquad+ 4a^4\big(58+45x_1^2+220x_2+12x_1(9-2x_2)\big)\big\}{\bf I}_4 \nn\\
 &&\quad+2\big\{1+306x_1^2+528x_2+672x_2^2+216x_3+648x_2x_3+648x_3^2+12x_1(16+75x_2+27x_3) \nn\\
 &&\qquad- 3a\big(10+405x_1^2+12x_1(24+97x_2+36x_3)+4(190x_2+197x_2^2+90x_3-432x_3^2)\big) \nn\\
 &&\qquad+ a^2\big(133+1818x_1^2+4008x_2^2+864x_3+72x_2(49+72x_3)+48x_1(31+111x_2+81x_3)\big) \nn\\
 &&\qquad- 12a^3\big(15+99x_1^2+220x_2+148x_2^2+12x_1(9+19x_2)\big) \nn\\
 &&\qquad+ 2a^4\big(38+261x_1^2+432x_2+432x_2^2+24x_1(10+27x_2)\big)\big\}m^2{\bf I}_1 \nn\\
 &&\quad+ 2\big\{187-45x_1^2+24x_2-156x_2^2+12x_1(31-15x_2) \nn\\
 &&\qquad- 6a\big(125-27x_1^2-4x_2-68x_2^2+6x_1(39-14x_2)\big) \nn\\
 &&\qquad+ a^2\big(1111-234x_1^2-144x_2-240x_2^2+24x_1(83-12x_2)\big) \nn\\
 &&\qquad- 48a^3\big(15-9x_1^2-2x_2-2x_2^2+3x_1(9-2x_2)\big) \nn\\
 &&\qquad+ 4a^4(43+84x_1-18x_1^2)\big\}m^2{\bf I}_2 + {\mathcal O}(k^4)\Big] + {\rm finite} \,. \nn
\eea
From (\ref{eqn:lmgi}), we can extract 
\be
{\bf B}_{\rm div} = -3(a)\,,\;\; {\bf \Pi}_{\rm div} = 12(b) + 18(c) + 18(d) \,,
\ee 
and the corresponding $b_i$'s can be obtained straightforwardly.


\begin{thebibliography}{99}

\bibitem{Fierz:1939ix}
  M.~Fierz and W.~Pauli,
  Proc.\ Roy.\ Soc.\ Lond.\  A {\bf 173}, 211 (1939).

\bibitem{Boulware:1973my}
  D.~G.~Boulware and S.~Deser,
  Phys.\ Rev.\  D {\bf 6}, 3368 (1972).

\bibitem{Creminelli:2005qk}
  P.~Creminelli, A.~Nicolis, M.~Papucci and E.~Trincherini,
  JHEP {\bf 0509}, 003 (2005)
  [arXiv:hep-th/0505147]; \\
  C.~de Rham and G.~Gabadadze,
  Phys.\ Lett.\  B {\bf 693}, 334 (2010)
  [arXiv:1006.4367 [hep-th]]; \\
  C.~de Rham and G.~Gabadadze,
  Phys.\ Rev.\  D {\bf 82}, 044020 (2010)
  [arXiv:1007.0443 [hep-th]]; \\
  L.~Berezhiani and M.~Mirbabayi,
  arXiv:1010.3288 [hep-th]; \\
  L.~Alberte, A.~H.~Chamseddine and V.~Mukhanov,
  arXiv:1011.0183 [hep-th]; \\
  C.~de Rham, G.~Gabadadze and A.~J.~Tolley,
  arXiv:1011.1232 [hep-th].

\bibitem{Park:2010rp}
  M.~Park,
  Class.\ Quant.\ Grav.\  {\bf 28}, 105012 (2011).
  [arXiv:1009.4369 [hep-th]]; \\
  D.~Metaxas,
  arXiv:1010.0246 [hep-th].

\bibitem{Stelle:1976gc}
  K.~S.~Stelle,
  Phys.\ Rev.\  {\bf D16}, 953-969 (1977).

\bibitem{vanDam:1970vg}
  H.~van Dam and M.~J.~G.~Veltman,
  Nucl.\ Phys.\  B {\bf 22}, 397 (1970). 

\bibitem{ArkaniHamed:2002fu}
  N.~Arkani-Hamed, S.~Dimopoulos, G.~Dvali and G.~Gabadadze,
  arXiv:hep-th/0209227.

\bibitem{Hawking:2001yt}
  S.~W.~Hawking and T.~Hertog,
  Phys.\ Rev.\  D {\bf 65}, 103515 (2002)
  [arXiv:hep-th/0107088]; \\
  C.~M.~Bender, S.~F.~Brandt, J.~H.~Chen and Q.~h.~Wang,
  Phys.\ Rev.\  D {\bf 71}, 025014 (2005)
  [arXiv:hep-th/0411064].

\bibitem{Lee:1969fy}
  T.~D.~Lee and G.~C.~Wick,
  Nucl.\ Phys.\  B {\bf 9}, 209 (1969).

\bibitem{Tomboulis:1977jk}
  E.~Tomboulis,
  Phys.\ Lett.\  {\bf B70}, 361 (1977).


\bibitem{Grinstein:2007mp}
  B.~Grinstein, D.~O'Connell and M.~B.~Wise,
  Phys.\ Rev.\  D {\bf 77}, 025012 (2008)
  [arXiv:0704.1845 [hep-ph]].

\bibitem{Cutkosky:1969fq}
  R.~E.~Cutkosky, P.~V.~Landshoff, D.~I.~Olive and J.~C.~Polkinghorne,
  Nucl.\ Phys.\  B {\bf 12}, 281 (1969); \\
  S.~Coleman,
  ``Acausality,'' In {\it Erice 1969, Ettore Majorana School On Subnuclear Phenomena}, New York 1970, 282-327.

\bibitem{Jansen:1993jj}
  K.~Jansen, J.~Kuti and C.~Liu,
  Phys.\ Lett.\  B {\bf 309}, 119 (1993)
  [arXiv:hep-lat/9305003]; \\
  A.~van Tonder,
  arXiv:0810.1928 [hep-th]; \\
  B.~Grinstein, D.~O'Connell and M.~B.~Wise,
  Phys.\ Rev.\  D {\bf 79}, 105019 (2009)
  [arXiv:0805.2156 [hep-th]].

\bibitem{Donoghue:1994dn}
  J.~F.~Donoghue,
  Phys.\ Rev.\  D {\bf 50}, 3874 (1994)
  [arXiv:gr-qc/9405057].

\bibitem{gvertices}
  B.~S.~DeWitt,
  Phys.\ Rev.\  {\bf 162}, 1239 (1967); \\
  S.~Sannan,
  Phys.\ Rev.\  D {\bf 34}, 1749 (1986).

\bibitem{vainshtein}
  A.~I.~Vainshtein,
  Phys.\ Lett.\  B {\bf 39}, 393 (1972).

\bibitem{Khoury:2003aq}
  D.~F.~Mota and J.~D.~Barrow,
  Phys.\ Lett.\  B {\bf 581}, 141 (2004)
  [arXiv:astro-ph/0306047]; \\
  J.~Khoury and A.~Weltman,
  Phys.\ Rev.\ Lett.\  {\bf 93}, 171104 (2004)
  [arXiv:astro-ph/0309300].

\bibitem{Higuchi:1986py}
  A.~Higuchi,
  Nucl.\ Phys.\  {\bf B282}, 397 (1987).

\bibitem{Grisa:2009yy}
  L.~Grisa, L.~Sorbo,
  Phys.\ Lett.\  {\bf B686}, 273-278 (2010).
  [arXiv:0905.3391 [hep-th]]; \\
  F.~Berkhahn, D.~D.~Dietrich, S.~Hofmann,
  JCAP {\bf 1011}, 018 (2010).
  [arXiv:1008.0644 [hep-th]].


\end{thebibliography}
\end{document}